\documentstyle[preprint,aps]{revtex}
\begin{document}

\newcommand{\btu}{\bigtriangledown}
\newcommand{\bge}{\begin{equation}}
\newcommand{\ege}{\end{equation}}
\newcommand{\bga}{\begin{eqnarray}} 
\newcommand{\ega}{\end{eqnarray}}
\newcommand{\nnu}{\nonumber}
\newcommand{\m}{{\cal M}}
\newcommand{\ep}{\epsilon}
\newcommand{\plb}{ Phys. Lett. {\bf B}}
\newcommand{\npb}{ Nucl. Phys. {\bf B}}

\draft
\preprint{\vbox{\hbox{IP/BBSR/96-18}\hbox{March 1996}\hbox{hep-th/9603110}} }
\title {\bf  S-duality and Canonical Transformations in String Theory}

\author{ Jnanadeva Maharana\footnote{Jawaharlal Nehru Fellow } 
and Harvendra Singh\footnote{e-mail: maharana@iopb.ernet.in,
hsingh@iopb.ernet.in}} 

\address{ Institute of Physics, Sachivalaya Marg, Bhubaneswar 751 005, India}

\maketitle

\begin{abstract} 

The symmetries of the tree level string effective action are
discussed. An appropriate effective action is constructed 
starting from the manifestly SL(2,R) invarint form of string
effective action introduced by Schwarz and Sen. 
The conserved charges are derived and generators of
infinitesimal transformations are obtained in the Hamiltonian
formalism. Some interesting consequences of the canonical
transformations are explored.
\end{abstract}
\vskip 1in

\narrowtext

\newpage

It is now recognized that dualities play a crucial role in our
understandings of string theory and field theory. The target
space duality \cite{1,ag}, termed as T-duality, is known to be a
symmetry of the string effective action and holds good order by
order in string perturbation theory. The
S-duality allows us to relate the strong and weak coupling
phases \cite{sen}. The Olive-Montonen conjecture \cite{2} implies
that the weak coupling and strong coupling regimes of
gauge theories are connected through the duality
transformation and these results were generalized to supersymmetric
cases subsequently\cite{osb}. Recently, considerable attention has been
focussed to understand various salient features of SUSY
Yang-Mills theories in this context \cite{4}.

It is well known that dilaton has a special status in string
theory. On the one hand it appears as a massless excitation of
bosonic as well as superstring like other massless excitations such as
graviton, antisymmetric tensor and gauge bosons and on the other
hand the vacuum 
expectation value of this field is the loop expansion parameter
of string theory and $e^\phi$ is identified as the string
coupling constant. We recall that, in four space-time
dimensions, the dual of the antisymmetric tensor field is a
pseudoscalar field, $\lambda_1$, identified with axion. Thus we can define a
complex field $\lambda=\lambda_1+i
\,\lambda_2,\,\,\,\lambda_2=e^{-\phi}$, and under an SL(2,R) 
transformation
$$\lambda\to \lambda '={a\,\lambda+b\over c\,\lambda
+d},\,\>\>\>\,\,\,a\,d-b\,c=1.$$ 
In fact SL(2,R) breaks to discrete S-duality subgroup SL(2,Z)
and this is expected to be an 
exact symmetry of string theory. Indeed, this symmetry can only
be tested nonperturbatively. It is well known  that the equations of
motion of the string effective action, with some constraints
\footnote{ The equations of motion do not remain invariant under
S-duality in presence of cosmological constant term in the
string effective action in certain cases as has been discussed
in \cite{kms}},
are invariant under S-duality transformations whereas the action
is invariant under T-duality and/or O(d,d) transformations.

There have been efforts to study the duality properties of
$\sigma$-model string world sheet action through canonical
transformations \cite{6,gv,7,can}. 
It is worth mentioning that the local symmetry properties of
string theories were studied by implementing suitable canonical
transformations in the frame work of BRST Hamiltonian path
integral formalism \cite{ven,jm}.
The purpose of this article is to investigate the symmetries of
the four-dimensional heterotic string effective action and
construct generators of the S-duality transformations and explore
some interesting properties of the action. 

In what follows, we consider a four-dimensional string effective
action derived by dimensionally reducing 10-dimensional
heterotic string effective action on six-dimensional torus,
$T^6$ \cite{9}:

\bga
S = \int d^4 x \sqrt{-G}\,\,e^{-\phi} \big ( R_G +
G^{\mu\nu}\partial_\mu\phi\partial_\nu\phi 
&&\>+ \>{1\over8} Tr \partial_\mu\, M^{-1}\partial^\mu M \nnu\\
&&- {1\over4} {\cal F}^i_{\mu\nu}\, ( M^{-1} )_{i\,j}\,{\cal
F}^{j\,\mu\nu} -{1\over12} H_{\mu\nu\lambda}\,H^{\mu\nu\lambda}
\big ).
\label{3}
\ega
$M$ is defined in terms of scalar fields arising from
dimensional reduction of metric, antisymmetric tensor and 16
gauge fields ( belonging to the Cartan subalgebra of the gauge
field sector of heterotic string theory ). Note that $M$
parametrises the coset $O(6,22)\over O(6)\times )(22)$ and $\phi$
is the shifted dilaton. The field strengths  are 

\bga
H_{\mu\nu\lambda}&=& \partial_\mu B_{\nu\lambda} -{1\over2}{\cal
A}_\mu^i {\cal L}_{i\,j} {\cal F}^j_{\nu\lambda} + cyclic\,\, perm.\nnu\\
{\cal F}^i_{\mu\nu}&=& \partial_\mu {\cal A}^i_\nu -
\partial_\nu{\cal A}^i_\mu \>, \,\,\,\,i=1,...,28,\nnu
\ega
${\cal A}_\mu^i$ are 28 gauge fields which
transform as vectors under O(6,22).
The action (\ref{3}) is manifestly invariant under O(6,22) global noncompact
transformations given by
 \bga
M&&\to \Omega \, M \, \Omega^T\;\; ,\nnu\\ 
\phi&&\to \phi,\,\,\, g_{\mu\nu}\to
g_{\mu\nu},\,\,\,B_{\mu\nu}\to B_{\mu\nu}, 
\>\>\> {\cal A}^i_\mu\to \Omega^i_j \,{\cal A}^j_\mu,\nnu\\
\Omega\,&&{\cal L}\,\Omega^T={\cal L}, \,\,\,\,{\cal
L}=\pmatrix{0&I_6&0\cr I_6&0&0\cr 0&0&I_{16}},
 \label{odd}
 \ega
${\cal L}$ is the O(d,d) metric, where $I_d$ is d-dimensional
identity matrix.  Note that the metric appearing in (\ref{3}) is the string
$\sigma$-model metric and scalar curvature, $R_G$, is computed
with respect to this metric.
Our goal is to study S-duality properties of this four
dimensional theory; for this purpose it is more convenient to go over
to the Einstein-frame metric
$$ G_{\mu\nu}\to g_{\mu \nu}=e^{-\phi} G_{\mu\nu}.$$
We mention in passing that equations of motion obtained from
the action ( derived from (\ref{3}) after above rescaling of the
metric ) are invariant under S-duality and that action itself
does not respect S-duality invariance. 
However, one can follow Schwarz and Sen\cite{ss} and reexpress
the action in manifestly SL(2,R) invariant form by introducing
an appropriate set of gauge fields. 
( From now on, we shall define the action in the Einstein frame
and $g_{\mu \nu}$ will stand for the Einstein metric. )

The action (\ref{3}) can be rewritten in manifestly SL(2,R)
invariant form as 
\bge
S=S_1 + S_2 + S_3 + S_4,
\label{act}
\ege
where $S_a,\,a=1,...,4$  terms have following contents,
\bga
S_1 =&& \int d^4 x \sqrt{- g} \,\big ( R 
+ {1\over8} \,Tr (\partial_\mu M^{-1}\partial^\mu M )\big), \nnu\\
S_2=&& {1\over 4}\int d^4x \sqrt{-g}\, tr(\partial_\mu {\cal
M}^{-1}\partial^\mu {\cal M} ), \nnu\\
S_3=&&-{1\over4}\int d^4x\sqrt{-g}  \,F^{(m, \alpha)}_{\mu\nu} {\hat G}_{m
n}\,{\cal M}^{-1}_{\alpha \beta} F^{(n, \beta)\,\mu\nu}, \nnu\\
S_4=&& - {1\over4} \int d^4x \sqrt{-g}\, F^{(m, \alpha)}_{\mu\nu} {\hat B}_{m
n}\,{\cal \eta}_{\alpha \beta}{\tilde F}^{(n, \beta)\,\mu\nu}.
\label{4}
\ega
Note that here and every where $g_{\mu\nu}$ denotes
four-dimensional Einstein metric;
space-time indeces $ \mu,\nu= 0,..., 3 $, internal indices
$ m,n=1,...,6$ and SL(2,R) indices $(\alpha,\beta)$ run over 
(1, 2). The moduli matrix $M$ parametrizes the coset $O(6,6)\over
O(6)\times O(6)$, matrix ${\cal M}$ parametrizes the coset
$SL(2,R)\over U(1)$.  $F_{\mu\nu}^{(m,\alpha)}$ represents
a SL(2,R) covariant set of gauge fields which come
from the compactification of the 
heterotic string on the torus and the auxiliary $U(1)$ fields.
These set of gauge fields transform as a vector under  SL(2,R) 
transformations. We note that the 16 abelian gauge fields of
10-dimensional heterotic string effective action have been set
to zero in the above action (\ref{act}) for the sake of
convenience. Now, the action (\ref{act}) does not have manifest
O(d,d) invariance; however, equations of motion still exhibit
O(d,d) invariance. We refer the reader to \cite{ss} for details
of the construction of the action (\ref{act}). The gauge field
strength and the matrix ${\cal M}$ are defined below,
\bga
&&F^{(m,\alpha)}_{\mu\nu}= \partial_\mu { A}^{(m,\alpha)}_\nu -
\partial_\nu{ A}^{(m,\alpha)}_\mu , \,\,\,\,\, {\tilde
F}^{\mu\nu\,(m,\alpha)}= {1\over2 \sqrt{-g}}
\ep^{\mu\nu\lambda\sigma}\,F^{(m,\alpha)}_{\lambda\sigma}\\ 
&& {\cal M}= {1\over\lambda_2}\pmatrix{
1 & \lambda_1 \cr \lambda_1 & |\lambda|^2},\,\,\,\,\,\lambda =
\lambda_1 + i \lambda_2, 
\label{5}
\ega
where ${\cal M}$ is a SL(2,R) matrix satisfying the constraints,
 
\bge
\m^T=\m, \,\,\,{\cal M}^T \eta {\cal M} = \eta,\,\,\,
\eta=\pmatrix{0&1\cr -1& 0},
\label{con}
\ege
$\eta$ being the SL(2,R) metric.
As was shown in \cite{ss} the above action has explicit
invariance under the following SL(2,R) transformatioms,
\bge
\m \to \omega^T\,\m\, \,\omega,\,\,\,
A_\mu \to \omega^T \, A_\mu,
\label{8}
\ege
where $\omega\in SL(2,R)$ matrix and satisfies $\omega^T \eta \,
\omega =\eta$. It was noted earliar that the invariance of the action
is achieved by  doubling the number of gauge fields. 

In what folllows, we present the transformation properties of
the fields under infinitesimal SL(2,R) group. 
These transformations are
\bga
&&\omega = 1 + \ep,\nnu\\
&&\delta\m = \ep^T \, \m + \m \, \ep,\hspace{1cm}
\delta A_\mu = \ep^T \, A_\mu,
\ega
where infinitesimal $2\times 2$ matrix $\ep$ satisfies the constraint $\ep^T
= \eta \, \ep \, \eta$. All other fields remain invariant under
these transformations. If $\,\Sigma^i$ are the generators of
SL(2,R) then we can write,
\bge
\ep = \alpha^i \, \Sigma^i,\,\,\,\,i=1,...,3,
\ege
$\alpha^i$'s being infinitesimal constant parameters.
 $\Sigma^i$ have following $2\times 2$ matrix representation; one can
use the  combination ( $\sigma_3 , \sigma_1, i\sigma_2$ );

\bga
 \Sigma^1=\pmatrix{ 1&0\cr0&-1},
\Sigma^2=\pmatrix{0&1\cr1&0},
\Sigma^3=\pmatrix{0&1\cr-1&0},
\ega
which satisfy the  algebra:
$$ [\Sigma^i, \Sigma^j] = 2 f^k_{i j}\, \Sigma^k, \,\,\,\,( i=1,2,3 )$$
and
$$ Tr( \Sigma^i \Sigma^j ) = 2 h^{i j} $$
where
$h_{i j}= diag( 1, 1, -1 )$ and $(\Sigma^1)^2=1,(\Sigma^2)^2=1,
(\Sigma^3)^2=-1$.
The $f^k_{i j}$'s are the structure constants satisfying
antisymmetric property $f^k_{i j} = -f^k_{j i} $. The
nonvanishing ones  are 
$$ f^3_{1 2}=1, f^1_{2 3}=-1, f^2_{3 1}=-1 .$$
These matrices also satisfy following relation, 
$$\Sigma^{i T} \eta \Sigma^j =-h^{i j}\,\eta.$$

The action (\ref{act}) being invariant under finite SL(2,R)
transformations (\ref{8}), also
respects invariance under  infinitesimal ones.
Consequently, we are in a position to reveal the underlying
conservation laws. Let us proceed to construct ``generating
functions'' of the infinitesimal $SL(2,R) $ transformations. 

We note that $S_1$, appearing in (\ref{act}), remains unaffected
by SL(2,R) transformations since it involves the metric $g_{\mu
\nu}$. Therefore from now on, we shall not explicitly mention the
action $S_1$ in our discussions; although the full action
contains contributions of $S_1$.
We now examine the variations of $S_2$, $S_3$ and $S_4$ under (9);

Notice that any matrix $P$ satisfying the constraint $P^T \,\eta\, P=\eta$
can be written in terms of the generators $\Sigma^i$ as
follows, 
\bge
P = p^0 \, I_2 + p^i\, \Sigma^i,\,\,\,i=1,...3.
\label{12}
\ege
Note that there are only three independent parameters in
(\ref{12}) due to the constraint on $P$. While an $SL(2,R)\over
U(1)$ matrix ${\bar \m}=\eta\,\m$ can be expressed as
\bge
{\bar \m}= m^i\,\Sigma^i
\ege
where only two of the $m^i$'s  are independent due to the constraint
$m_3^2-m_2^2-m_1^2=1$ which follows from the constraints on
$\m$ in (\ref{con}). Thus we can rewrite the action  $S_2$  as 
\bge
S_2= -{1\over2} \int d^4 x \sqrt{-g} \, h^{i j} \,\partial_\mu m_i
\,\partial^\mu m_j,
\label{16}
\ege
where the metric $h = {\rm diag}( 1, 1 ,-1)$,~ $m_1={
\lambda_1\over\lambda_2}, ~ m_2={|\lambda|^2 
-1\over2\lambda_2}\,{\rm and}\,~ m_3={|\lambda|^2
+1\over2\lambda_2}$ \footnote{ we already know that $m_i$'s depend
on two independent fields $\lambda_1$ and $\lambda_2$. It is a
reflection of the fact that the matrix $\m$ parametrizes the
coset $SL(2,R)\over U(1)$.}.
Correspondingly, one can deduce that the infinitesimal SL(2,R) transformation
of $m_i$ is
\bge 
\delta m_k= f^k_{ i j} \, \alpha_i \, m_j.
\ege
Explicitly
\bga
\delta \pmatrix{m_1\cr m_2\cr m_3}&&= \pmatrix{0 & -\alpha_3 &
\alpha_2\cr \alpha_3 &0 & -\alpha_1\cr \alpha_2 & -\alpha_1 & 0}
\pmatrix{m_1\cr m_2 \cr m_3} \nnu\\
&&= - \sum_{i=1}^3 \alpha_i \, \Gamma_i \,  \pmatrix{m_1\cr m_2 \cr m_3},
\label{161}
\ega
where the representation for $\{\Gamma^i\}$ is; 
\bga
\Gamma^1=\pmatrix{0&0&0\cr0&0&1\cr 0&1&0},
\Gamma^2=\pmatrix{0&0&-1\cr0&0&0\cr -1&0&0},
\Gamma^3=\pmatrix{0&1&0\cr-1&0&0\cr 0&0&0},
\ega
which are traceless and satisfy the following algebra:
$$[\Gamma^i, \Gamma^j]= f^k_{i j} \Gamma^k $$ and
$$Tr[\Gamma^i\,\Gamma^j]= 2 h^{i j},$$ 
$$\Gamma^{i T}= h\,\Gamma^i\,h .$$

Action in (\ref{16}) is invariant under (\ref{161}).

Let us now turn to the action $S_3$, we can write
the action in the following form 
\bge
S_3= \int d^4x \sqrt{-g} \,m_i \,\,K^i,
\ege
where 
\bge
K^i  = {1\over 4}\,F^{(m, \alpha)}_{\mu\nu} {\hat G}_{m
n}\,(\Sigma^i \,\eta)_{\alpha \beta}\, F^{(n, \beta)\,\mu\nu}.
\ege
Using the SL(2,R) transformations for vector fields $A^{(m,
\alpha)}_\mu$ we derive the transformation on $K^i$ to be the following;
\bga
\delta K^i= - \{ \sum_k \alpha_k \, ( h \,\Gamma_k\, h)\}_{i j} \, K^j.
\ega
Note that $ K^i = h^{i j} K_j$. Then the covariant $K_i$ will transform as
\bge
\delta K_i= - (\sum_k \alpha_k \,\Gamma_k)_{i j} \, K_j.
\label{21}
\ege

Therefore, we can rewrite
\bge
S_2 + S_3= \int d^4x \sqrt{-g}\left( 
-{1\over2} \, h^{i j} \,\partial_\mu m_i \,\partial^\mu m_j
+ m_i\,K_j \, h^{i j}\right).
\label{act1}
\ege
which is invariant under (\ref{161}) and (\ref{21}). It can be
checked that $S_4$ also remain invariant. From here the
canonical momentum conjugate to $m_i$ is 
$\pi^i= \partial^0 \, m^i,$ which transforms as
\bge 
\delta\pi^i = ( \alpha\,.\Gamma^T )_{i j} \, \pi^j.
\label{22}
\ege
We remind the reader that a classical mechanical system in the
Hamiltonian phase space formalism is described by a set of
generalised coordinates $\{ q_i\}$ and conjugate momenta $\{
p^i\}$. Under infinitesimal canonical transformations they
transform as 
\bga
q'_i= q_i -\,\alpha\,. \, \left({\partial
\Phi_{(q)}(q,p)\over\partial p^i}\right), \nnu\\
p'^i= p^i +\,\alpha\,. \, \left({\partial
\Phi_{(q)}(q,p)\over\partial q_i}\right),
\label{24}
\ega
where $\Phi_{(q)}$ is the generator of the transformation. 
Thus for the case of SL(2,R), we can construct the generators of
infinitesimal transformations to be  
\bge
\Phi_{(m)}^k(m, \pi)= \int \pi^j \,(\Gamma^k)_{j i} \, m_i.
\ege
Similarly, the canonical momentum conjugate to $A^{(m,
\alpha)}_\mu$ is ( here we suppress the indices for
convenience) 
$$ \pi^{\mu}={\hat G} \m^{-1}\, F^{0 \mu} + {\hat B}
\,\eta \, {\tilde F}^{0 \mu},$$ and it  transforms as
\bge
\delta \pi^\mu = - \alpha\,.\,\Sigma \,\pi^\mu.
\ege
Therefore, the corresponding generating function in the gauge field sector
is given by
\bge
\Phi_{(A)}^i( A_\mu, \pi^\mu ) = - \int \pi^{\mu\,(m, \alpha)} \,
(\Sigma^i)_{\alpha \beta} \, A^{(m, \beta)}_\mu.
\ege
We can now write down the complete generating function for
canonical transformation in the phase space of the full theory
to be the sum of above two generating functions. That is
\bga
\Phi^i=\Phi^i_{(m)}+\Phi^i_{(A)}.
\ega

Now, we shall obtain conserved charges associated with the
infinitesimal SL(2,R) transformations. 
 We obtain the conserved current for the scalars $m_i$ to be
\bge
J_{(m)}^{k, \mu}=  m_i\,(\Gamma^k)_{i j} \,\partial^\mu m_j
\ege
and that for gauge fields $A_\mu$ is
\bge
J_{(A)}^{i, \mu} = A_\nu \,\Sigma^{i T} \, {\hat G}\, \m^{-1} \,F^{\mu \nu}.
\ege
Corresponding Noether charges are
\bga
&&Q_{(m)}^a=\int d^3x \sqrt{-g}\, m_i \Gamma^a_{i j} \pi^j,\nnu\\
&&Q_{(A)}^a=\int d^3x \sqrt{-g}\, A_\nu \,\Sigma^{a T} \,\pi^\nu,
\ega
respectively. They satisfy the following algebra,
\bga
&&\{ Q_{(m)}^a, m_i\}=  \Gamma^a _{i j} \, m_j \equiv (\delta m_i)^a,\nnu\\
&&\{ Q_{(A)}^a, A_\mu^{(m, \alpha)}\}= \Sigma^a_{\alpha \beta}
\, A_\mu^{(m, \beta)}\equiv(\delta A_\mu^{(m,\alpha)})^a.
\ega

At this point we can discuss some properties of the generating
functional which follow 
from the invariance of the classical path integral under the canonical
transformations in the Hamiltonian approach.
One can write down the generating functional for correlation
functions, $Z[\zeta_i,J^\mu]$, in the
phase space,

\bge
Z[\zeta,J]= \int {\cal D}[m_i,\pi_i,A_\mu,\pi_\mu]
\hskip .5cm e^{\{i \,S_H[m,A,\pi] + ({\rm source\>terms})\}} 
\ege
where source terms are
\bge
 {\rm source\> terms}= i \int ( m^i \zeta_i + A_\mu^{(m,\alpha)}
J^\mu_{(m, \alpha)}),
\ege
with $\zeta_i$ and $J^\mu_{(m \alpha)}$ being the classical
sources and summation over repeated indices is understood.
${\cal D}[m_i,\pi_i,A_\mu,\pi_\mu]$ collectively stands for the
Hamiltonian phase space measure and $S_H$ is the Hamiltonian
action. 

Now we adopt the procedure of Veneziano, further elaborated by
Maharana and Veneziano, to exhibit some interesting properties of
the partition function \cite{ven,jm}. Note that when we implement
infinitesimal canonical transformation, the variables in the
Hamiltonian phase space change according to eq.(24). The
Hamiltonian action $S_H$ and field variables appearing in source terms 
(34) transform accordingly which can be
compensated by a shift in the sources. On the other hand,
the phase space measure remains invariant, modulo anomaly terms.
In ref. \cite{ven} and \cite{jm}, this property was exploited to
derive Ward identities. 

It is easy to see that for the problem
at hand the following shifts of the sources
\bge
 \delta \zeta_i = -\{\alpha_k\,\Gamma_k\}_{i j}\zeta_j,\hspace{1cm} \delta
J^\mu_{(m,\alpha)}= \{\alpha_k \Sigma_k\}_{\alpha \beta}
J^\mu_{(m,\beta)} .
\ege
enable us to derive a relation which is satisfied by the 
generating functional
\bge
Z[\zeta,J] = Z[\zeta+\delta\zeta , J + \delta J].
\label{gf}
\ege
Now following the arguments of Maharana and Veneziano we are led to
\bge
\int d^4x \left(- {\delta Z\over \delta \zeta_i(x) }\,(\Gamma^k)_{i j}
\zeta_j(x) + {\delta Z \over \delta J^\mu_{(m\alpha)}(x)}
(\Sigma^k)_{\alpha\beta} J^\mu_{(m \beta)}(x)\right) \,\,\alpha^k=0.
\label{wi}
\ege
We mention in passing that the Hamiltonian approach is an
elegant way to derive identities like (\ref{wi}) which can also be
obtained in the Lagrangian formulation\cite{lee}.
We recall that an equation like (\ref{wi}) was the starting
point of the derivation of gravitational and gauge Ward
identities in ref. \cite{ven} where the infinitesimal parameter
was a local one. For the case at hand we are dealing with global
symmetries and the infinitesimal parameters appearing in
(\ref{wi}) are independent of spacetime. Notice, however, that
the relation holds for arbitrary infinitesimal parameters and
therefore, we can differentiate this equation with respect to
$\alpha^k$ and then set $\alpha^k=0$. Thus we arrive at 

\bge
\int \left(- {\delta Z\over \delta \zeta_i }\,(\Gamma^k)_{i j}
\zeta_j + {\delta Z \over \delta J^\mu_{(m\alpha)}}
(\Sigma^k)_{\alpha\beta} J^\mu_{(m \beta)}\right)=0.
\label{wii}
\ege

Now we can functionally differentiate (\ref{wii}) with respect
to the sources $\zeta_i$ and $J^\mu_{(m,\alpha)}$ several times
and then set these sources to zero. In this process, we shall
obtain interesting relations involving correlation functions
which will be analogous to the Ward identities derived in
\cite{ven}. We mention that these are not the Ward identities
one derives for local symmetries; but our relations are obtained
in the context of global symmetries.

To summarise, we have studied the SL(2,R), identified as the
S-duality group, transformation properties of the string
effective action. We start from an effective action introduced
in \cite{ss} which contains an appropriate set  of gauge fields to
be manifestly SL(2,R) invariant. The actions $S_2$ and $S_3$ are
reexpressed in a suitable form so that we can exploit the
symmetry properties to derive eq.(\ref{wii}).
The conserved currents are obtained and the
generators of canonical transformation responsible for
infinitesimal transformations are identified. 
We obtain a set of relations for correlation functions by
exploiting the symmetry properties of the effective action under
canonical transformations.


\begin{references}

\bibitem{1} For review see: A. Giveon, M. Porrati and E.
Ravinovici Phys. Rep.{\bf 244C}, 77(1994).

\bibitem{ag} E. Alvarez, L. Alvarez-Gaume and Y. Lozano, Nucl.
Phys. {\bf B41}, 1 (1995), (hep-th/9410237); E. Alvarez, L.
Alvarez-Gaume, J.L.F. Barbon and Y. Lozano, Nucl. Phys. {\bf
B415}, 71(1994); L. Alvarez-Gaume, trieste lectures.

\bibitem{sen} For review and references see: A. Sen, Int. J.
Mod. Phys. {\bf A9}, 3707 (1994). 

\bibitem{2}  C. Montonen and D. Olive, Phys. Lett. {\bf B72}, 117 (1977); P.
Goddard, J. Nyuts and D. Olive, Nucl. Phys. {\bf B125}, 1
(1977).
\bibitem{osb} H. Osborn, Phys. Lett. {\bf B83}, 321(1979); E.
Witten and D.Olive, Phys. Lett. {\bf B78}, 97 (1978). 

\bibitem{4} C. Vafa and E. Witten, preprint HUTP-94-A017; 
N. Seiberg and E. Witten, Nucl. Phys. {\bf B426}, 19 (1994); 
Nucl. Phys. {\bf B431}, 484 (1994); A. Ceresole, R.
D'Auria and S. Ferrara, Phys. Lett. {\bf B339}, 71 (1994);
M. Bershadsky, A. Johansen, V. Sadov and C. Vafa,
preprint HUTP-95-A004.

\bibitem{kms} S. Kar, J. Maharana and H. Singh, Phys. Lett.
B ( in press ), hep-th/9507063; J. Maharana and H. Singh, Phys.
Lett. {\bf B368} (1996) 64 (hep-th/9506213).

\bibitem{6} A. Giveon, E. Rabinovici and G. Veneziano, Nucl.
Phys. {\bf B}322, 167 (1989);
K.A. Meissner and G. Veneziano, \plb267, 33(1991)
\bibitem{gv} G. Veneziano, Phys. Lett. \plb285, 287(1991).

\bibitem{7} E. Alvarez, L. Alvarez-Gaume and Y. Lozano, Phys.
Lett. B336 (1994) 183.

\bibitem{can} Y. Lozano, {\it Nonabelian Duality and Canonical
Transformations}, PUPT-1532, hep-th/9503045;  {\it S-duality in Gauge
Theories as Canonical Transformation}, PUPT-1552,
hep-th/9508021; A. A. Kehagias, {\it A canonical approach to
S-duality in Abelian Gauge Theories}, NTUA-95/51, hep-th/9508159;
T. Curtright and C. Zachos, {\it Canonical Nonabelian Dual
Transformations in Superymmetric Field Theories},
hep-th/9502126. 

\bibitem{ven} G. Veneziano, Phys. Lett. B167 (1986) 388; J.
Maharana and G. Veneziano, Phys. Lett. B169 (1986) 177; Nucl.
Phys. B243 (1987) 126.

\bibitem{jm} J. Maharana, Phys. Lett. B211 (1988) 431.

\bibitem{9} J. Maharana and J. Schwarz, Nucl. Phys. {\bf B390},
4 (1993); S. Hassan and A. Sen, Nucl. Phys. {\bf B375}, 103 (1993).

\bibitem{ss} J. Schwarz and A. Sen, Phys. Lett. {\bf B312}, 105 (1993), Nucl.
Phys. {\bf B411}, 35 (1994).

\bibitem{lee} B. W. Lee, In {\it Methods of Field Theory, Les
Houches 1975},  eds. R. Balian and J. Zinn-Justin ( North Holland,
1976 ).



\end{references}
\end{document}